\documentclass[longauth]{aa}

\def\J1826{HESS\,J1826$-$130}
\usepackage{amsmath}
\usepackage{microtype}
\usepackage{lineno}
\usepackage[varg]{txfonts}
\usepackage{graphicx}
\usepackage{epstopdf}
\usepackage{url}
\usepackage{natbib}
\usepackage{xcolor}
\usepackage{multirow}
\usepackage{gensymb}
\usepackage{lscape}

\makeatletter
\renewcommand*\maketitle{%
  \thispagestyle{firstpage}
\begingroup
    \if@wideboxfn
    \setlength\bibindent{1.4\parindent}
    \else
    \setlength\bibindent{\parindent}
    \fi
    \renewcommand*\thefootnote{\@fnsymbol\c@footnote}%
    \renewcommand\@makefntext[1]{%
    \ifaa@longfn\hsize\textwidth\fi
    \noindent
    \hb@xt@\bibindent{\hss\@makefnmark\enspace}##1}
  \ifaa@twocolumn
  \begingroup
    \begin{aa@strip}
          \aa@maketitle
    \end{aa@strip}
    \@thanks	  	
  \endgroup
  \else
    \begingroup
      \let\thanks\footnote
      \aa@maketitle
    \endgroup
  \fi
\endgroup
  \setcounter{footnote}{0}%
}
\makeatother

\begin{document}

\title{An extreme particle accelerator in the Galactic plane: HESS\,J1826$-$130}

\makeatletter
\renewcommand*{\@fnsymbol}[1]{\ifcase#1\or*\or$\dagger$\or$\ddagger$\or**\or$\dagger\dagger$\or$\ddagger\ddagger$\fi}
\makeatother

\author{H.E.S.S. Collaboration
\and H.~Abdalla \inst{\ref{NWU}}
\and R.~Adam \inst{\ref{LLR}}
\and F.~Aharonian \inst{\ref{DIAS},\ref{MPIK},\ref{RAU}}
\and F.~Ait~Benkhali \inst{\ref{MPIK}}
\and E.O.~Ang\"uner\protect\footnotemark[1] \inst{\ref{CPPM}}
\and C.~Arcaro \inst{\ref{NWU}}
\and C.~Armand \inst{\ref{LAPP}}
\and T.~Armstrong \inst{\ref{Oxford}}
\and H.~Ashkar \inst{\ref{IRFU}}
\and M.~Backes \inst{\ref{UNAM},\ref{NWU}}
\and V.~Baghmanyan \inst{\ref{IFJPAN}}
\and V.~Barbosa~Martins \inst{\ref{DESY}}
\and A.~Barnacka \inst{\ref{UJK}}
\and M.~Barnard \inst{\ref{NWU}}
\and Y.~Becherini \inst{\ref{Linnaeus}}
\and D.~Berge \inst{\ref{DESY}}
\and K.~Bernl\"ohr \inst{\ref{MPIK}}
\and B.~Bi \inst{\ref{IAAT}}
\and M.~B\"ottcher \inst{\ref{NWU}}
\and C.~Boisson \inst{\ref{LUTH}}
\and J.~Bolmont \inst{\ref{LPNHE}}
\and M.~de~Bony~de~Lavergne \inst{\ref{LAPP}}
\and P.~Bordas\protect\footnotemark[1] \inst{\ref{CerrutiNowAt}}
\and M.~Breuhaus \inst{\ref{MPIK}}
\and F.~Brun \inst{\ref{IRFU}}
\and P.~Brun \inst{\ref{IRFU}}
\and M.~Bryan \inst{\ref{GRAPPA}}
\and M.~B\"{u}chele \inst{\ref{ECAP}}
\and T.~Bulik \inst{\ref{UWarsaw}}
\and T.~Bylund \inst{\ref{Linnaeus}}
\and S.~Caroff \inst{\ref{LPNHE}}
\and A.~Carosi \inst{\ref{LAPP}}
\and S.~Casanova\protect\footnotemark[1] \inst{\ref{IFJPAN},\ref{MPIK}}
\and T.~Chand \inst{\ref{NWU}}
\and S.~Chandra \inst{\ref{NWU}}
\and A.~Chen \inst{\ref{WITS}}
\and G.~Cotter \inst{\ref{Oxford}}
\and M.~Cury{\l}o \inst{\ref{UWarsaw}}
\and J.~Damascene~Mbarubucyeye \inst{\ref{DESY}}
\and I.D.~Davids \inst{\ref{UNAM}}
\and J.~Davies \inst{\ref{Oxford}}
\and C.~Deil \inst{\ref{MPIK}}
\and J.~Devin \inst{\ref{CENBG}}
\and P.~deWilt \inst{\ref{Adelaide}}
\and L.~Dirson \inst{\ref{HH}}
\and A.~Djannati-Ata\"i \inst{\ref{APC}}
\and A.~Dmytriiev \inst{\ref{LUTH}}
\and A.~Donath \inst{\ref{MPIK}}
\and V.~Doroshenko \inst{\ref{IAAT}}
\and C.~Duffy \inst{\ref{Leicester}}
\and J.~Dyks \inst{\ref{NCAC}}
\and K.~Egberts \inst{\ref{UP}}
\and F.~Eichhorn \inst{\ref{ECAP}}
\and S.~Einecke \inst{\ref{Adelaide}}
\and G.~Emery \inst{\ref{LPNHE}}
\and J.-P.~Ernenwein \inst{\ref{CPPM}}
\and K.~Feijen \inst{\ref{Adelaide}}
\and S.~Fegan \inst{\ref{LLR}}
\and A.~Fiasson \inst{\ref{LAPP}}
\and G.~Fichet~de~Clairfontaine \inst{\ref{LUTH}}
\and G.~Fontaine \inst{\ref{LLR}}
\and S.~Funk \inst{\ref{ECAP}}
\and M.~F\"u{\ss}ling \inst{\ref{DESY}}
\and S.~Gabici \inst{\ref{APC}}
\and Y.A.~Gallant \inst{\ref{LUPM}}
\and G.~Giavitto \inst{\ref{DESY}}
\and L.~Giunti \inst{\ref{APC}, \ref{IRFU}}
\and D.~Glawion \inst{\ref{LSW}}
\and J.F.~Glicenstein \inst{\ref{IRFU}}
\and D.~Gottschall \inst{\ref{IAAT}}
\and M.-H.~Grondin \inst{\ref{CENBG}}
\and J.~Hahn \inst{\ref{MPIK}}
\and M.~Haupt \inst{\ref{DESY}}
\and G.~Hermann \inst{\ref{MPIK}}
\and J.A.~Hinton \inst{\ref{MPIK}}
\and W.~Hofmann \inst{\ref{MPIK}}
\and C.~Hoischen \inst{\ref{UP}}
\and T.~L.~Holch \inst{\ref{HUB}}
\and M.~Holler \inst{\ref{LFUI}}
\and M.~H\"{o}rbe \inst{\ref{Oxford}}
\and D.~Horns \inst{\ref{HH}}
\and D.~Huber \inst{\ref{LFUI}}
\and M.~Jamrozy \inst{\ref{UJK}}
\and D.~Jankowsky \inst{\ref{ECAP}}
\and F.~Jankowsky \inst{\ref{LSW}}
\and A.~Jardin-Blicq \inst{\ref{MPIK}}
\and V.~Joshi \inst{\ref{ECAP}}
\and I.~Jung-Richardt \inst{\ref{ECAP}}
\and E.~Kasai \inst{\ref{UNAM}}
\and M.A.~Kastendieck \inst{\ref{HH}}
\and K.~Katarzy{\'n}ski \inst{\ref{NCUT}}
\and U.~Katz \inst{\ref{ECAP}}
\and D.~Khangulyan \inst{\ref{Rikkyo}}
\and B.~Kh\'elifi \inst{\ref{APC}}
\and S.~Klepser \inst{\ref{DESY}}
\and W.~Klu\'{z}niak \inst{\ref{NCAC}}
\and Nu.~Komin \inst{\ref{WITS}}
\and R.~Konno \inst{\ref{DESY}}
\and K.~Kosack \inst{\ref{IRFU}}
\and D.~Kostunin \inst{\ref{DESY}} 
\and M.~Kreter \inst{\ref{NWU}}
\and G.~Lamanna \inst{\ref{LAPP}}
\and A.~Lemi\`ere \inst{\ref{APC}}
\and M.~Lemoine-Goumard \inst{\ref{CENBG}}
\and J.-P.~Lenain \inst{\ref{LPNHE}}
\and C.~Levy \inst{\ref{LPNHE}}
\and T.~Lohse \inst{\ref{HUB}}
\and I.~Lypova \inst{\ref{DESY}}
\and J.~Mackey \inst{\ref{DIAS}}
\and J.~Majumdar \inst{\ref{DESY}}
\and D.~Malyshev \inst{\ref{IAAT}}
\and D.~Malyshev \inst{\ref{ECAP}}
\and V.~Marandon \inst{\ref{MPIK}}
\and P.~Marchegiani \inst{\ref{WITS}}
\and A.~Marcowith \inst{\ref{LUPM}}
\and A.~Mares \inst{\ref{CENBG}}
\and G.~Mart\'i-Devesa \inst{\ref{LFUI}}
\and R.~Marx \inst{\ref{LSW}, \ref{MPIK}}
\and G.~Maurin \inst{\ref{LAPP}}
\and P.J.~Meintjes \inst{\ref{UFS}}
\and M.~Meyer \inst{\ref{ECAP}}
\and A.~Mitchell \inst{\ref{MitchellNowAt}}
\and R.~Moderski \inst{\ref{NCAC}}
\and M.~Mohamed \inst{\ref{LSW}}
\and L.~Mohrmann \inst{\ref{ECAP}}
\and A.~Montanari \inst{\ref{IRFU}}
\and C.~Moore \inst{\ref{Leicester}}
\and P.~Morris \inst{\ref{Oxford}}
\and E.~Moulin \inst{\ref{IRFU}}
\and J.~Muller \inst{\ref{LLR}}
\and T.~Murach \inst{\ref{DESY}}
\and K.~Nakashima \inst{\ref{ECAP}}
\and A.~Nayerhoda \inst{\ref{IFJPAN}}
\and M.~de~Naurois \inst{\ref{LLR}}
\and H.~Ndiyavala  \inst{\ref{NWU}}
\and F.~Niederwanger \inst{\ref{LFUI}}
\and J.~Niemiec \inst{\ref{IFJPAN}}
\and L.~Oakes \inst{\ref{HUB}}
\and P.~O'Brien \inst{\ref{Leicester}}
\and H.~Odaka \inst{\ref{Tokyo}}
\and S.~Ohm \inst{\ref{DESY}}
\and L.~Olivera-Nieto \inst{\ref{MPIK}}
\and E.~de~Ona~Wilhelmi \inst{\ref{DESY}}
\and M.~Ostrowski \inst{\ref{UJK}}
\and I.~Oya\protect\footnotemark[1] \inst{\ref{DESY}}
\and M.~Panter \inst{\ref{MPIK}}
\and S.~Panny \inst{\ref{LFUI}}
\and R.D.~Parsons \inst{\ref{HUB}}
\and G.~Peron \inst{\ref{MPIK}}
\and B.~Peyaud \inst{\ref{IRFU}}
\and Q.~Piel \inst{\ref{LAPP}}
\and S.~Pita \inst{\ref{APC}}
\and V.~Poireau \inst{\ref{LAPP}}
\and A.~Priyana~Noel \inst{\ref{UJK}}
\and D.A.~Prokhorov \inst{\ref{GRAPPA}}
\and H.~Prokoph \inst{\ref{DESY}}
\and G.~P\"uhlhofer \inst{\ref{IAAT}}
\and M.~Punch \inst{\ref{APC},\ref{Linnaeus}}
\and A.~Quirrenbach \inst{\ref{LSW}}
\and S.~Raab \inst{\ref{ECAP}}
\and R.~Rauth \inst{\ref{LFUI}}
\and P.~Reichherzer \inst{\ref{IRFU}}
\and A.~Reimer \inst{\ref{LFUI}}
\and O.~Reimer \inst{\ref{LFUI}}
\and Q.~Remy \inst{\ref{MPIK}}
\and M.~Renaud \inst{\ref{LUPM}}
\and F.~Rieger \inst{\ref{MPIK}}
\and L.~Rinchiuso \inst{\ref{IRFU}}
\and C.~Romoli \inst{\ref{MPIK}}
\and G.~Rowell \inst{\ref{Adelaide}}
\and B.~Rudak \inst{\ref{NCAC}}
\and E.~Ruiz-Velasco \inst{\ref{MPIK}}
\and V.~Sahakian \inst{\ref{YPI}}
\and S.~Sailer \inst{\ref{MPIK}}
\and D.A.~Sanchez \inst{\ref{LAPP}}
\and A.~Santangelo \inst{\ref{IAAT}}
\and M.~Sasaki \inst{\ref{ECAP}}
\and M.~Scalici \inst{\ref{IAAT}}
\and F.~Sch\"ussler \inst{\ref{IRFU}}
\and H.M.~Schutte \inst{\ref{NWU}}
\and U.~Schwanke \inst{\ref{HUB}}
\and S.~Schwemmer \inst{\ref{LSW}}
\and M.~Seglar-Arroyo \inst{\ref{IRFU}}
\and M.~Senniappan \inst{\ref{Linnaeus}}
\and A.S.~Seyffert \inst{\ref{NWU}}
\and N.~Shafi \inst{\ref{WITS}}
\and K.~Shiningayamwe \inst{\ref{UNAM}}
\and R.~Simoni \inst{\ref{GRAPPA}}
\and A.~Sinha \inst{\ref{APC}}
\and H.~Sol \inst{\ref{LUTH}}
\and A.~Specovius \inst{\ref{ECAP}}
\and S.~Spencer \inst{\ref{Oxford}}
\and M.~Spir-Jacob \inst{\ref{APC}}
\and {\L.}~Stawarz \inst{\ref{UJK}}
\and L.~Sun \inst{\ref{GRAPPA}}
\and R.~Steenkamp \inst{\ref{UNAM}}
\and C.~Stegmann \inst{\ref{UP},\ref{DESY}}
\and S.~Steinmassl \inst{\ref{MPIK}}
\and C.~Steppa \inst{\ref{UP}}
\and T.~Takahashi  \inst{\ref{KAVLI}}
\and T.~Tavernier \inst{\ref{IRFU}}
\and A.M.~Taylor \inst{\ref{DESY}}
\and R.~Terrier \inst{\ref{APC}}
\and D.~Tiziani \inst{\ref{ECAP}}
\and M.~Tluczykont \inst{\ref{HH}}
\and L.~Tomankova \inst{\ref{ECAP}}
\and C.~Trichard \inst{\ref{LLR}}
\and M.~Tsirou \inst{\ref{LUPM}}
\and R.~Tuffs \inst{\ref{MPIK}}
\and Y.~Uchiyama \inst{\ref{Rikkyo}}
\and D.J.~van~der~Walt \inst{\ref{NWU}}
\and C.~van~Eldik \inst{\ref{ECAP}}
\and C.~van~Rensburg \inst{\ref{NWU}}
\and B.~van~Soelen \inst{\ref{UFS}}
\and G.~Vasileiadis \inst{\ref{LUPM}}
\and J.~Veh \inst{\ref{ECAP}}
\and C.~Venter \inst{\ref{NWU}}
\and P.~Vincent \inst{\ref{LPNHE}}
\and J.~Vink \inst{\ref{GRAPPA}}
\and H.J.~V\"olk \inst{\ref{MPIK}}
\and T.~Vuillaume \inst{\ref{LAPP}}
\and Z.~Wadiasingh \inst{\ref{NWU}}
\and S.J.~Wagner \inst{\ref{LSW}}
\and J.~Watson \inst{\ref{Oxford}}
\and F.~Werner \inst{\ref{MPIK}}
\and R.~White \inst{\ref{MPIK}}
\and A.~Wierzcholska \inst{\ref{IFJPAN},\ref{LSW}}
\and Yu Wun Wong \inst{\ref{ECAP}}
\and A.~Yusafzai \inst{\ref{ECAP}}
\and M.~Zacharias \inst{\ref{NWU},\ref{LUTH}}
\and R.~Zanin \inst{\ref{MPIK}}
\and D.~Zargaryan \inst{\ref{DIAS},\ref{RAU}}
\and A.A.~Zdziarski \inst{\ref{NCAC}}
\and A.~Zech \inst{\ref{LUTH}}
\and S.J.~Zhu \inst{\ref{DESY}}
\and A.~Ziegler\protect\footnotemark[1] \inst{\ref{ECAP}}
\and J.~Zorn \inst{\ref{MPIK}}
\and S.~Zouari \inst{\ref{APC}}
\and N.~\.Zywucka \inst{\ref{NWU}}
}

\institute{
Centre for Space Research, North-West University, Potchefstroom 2520, South Africa \label{NWU} \and 
Laboratoire Leprince-Ringuet, École Polytechnique, CNRS, Institut Polytechnique de Paris, F-91128 Palaiseau, France \label{LLR} \and
Dublin Institute for Advanced Studies, 31 Fitzwilliam Place, Dublin 2, Ireland \label{DIAS} \and 
Max-Planck-Institut f\"ur Kernphysik, P.O. Box 103980, D 69029 Heidelberg, Germany \label{MPIK} \and 
High Energy Astrophysics Laboratory, RAU,  123 Hovsep Emin St  Yerevan 0051, Armenia \label{RAU} \and
Aix Marseille Universit\'e, CNRS/IN2P3, CPPM, Marseille, France \label{CPPM} \and
Laboratoire d'Annecy de Physique des Particules, Univ. Grenoble Alpes, Univ. Savoie Mont Blanc, CNRS, LAPP, 74000 Annecy, France \label{LAPP} \and
University of Oxford, Department of Physics, Denys Wilkinson Building, Keble Road, Oxford OX1 3RH, UK \label{Oxford} \and
IRFU, CEA, Universit\'e Paris-Saclay, F-91191 Gif-sur-Yvette, France \label{IRFU} \and
University of Namibia, Department of Physics, Private Bag 13301, Windhoek 10005, Namibia \label{UNAM} \and
Instytut Fizyki J\c{a}drowej PAN, ul. Radzikowskiego 152, 31-342 Krak{\'o}w, Poland \label{IFJPAN} \and
DESY, D-15738 Zeuthen, Germany \label{DESY} \and
Obserwatorium Astronomiczne, Uniwersytet Jagiello{\'n}ski, ul. Orla 171, 30-244 Krak{\'o}w, Poland \label{UJK} \and
Department of Physics and Electrical Engineering, Linnaeus University,  351 95 V\"axj\"o, Sweden \label{Linnaeus} \and
Institut f\"ur Astronomie und Astrophysik, Universit\"at T\"ubingen, Sand 1, D 72076 T\"ubingen, Germany \label{IAAT} \and
Laboratoire Univers et Théories, Observatoire de Paris, Université PSL, CNRS, Université de Paris, 92190 Meudon, France \label{LUTH} \and
Sorbonne Universit\'e, Universit\'e Paris Diderot, Sorbonne Paris Cit\'e, CNRS/IN2P3, Laboratoire de Physique Nucl\'eaire et de Hautes Energies, LPNHE, 4 Place Jussieu, F-75252 Paris, France \label{LPNHE} \and
GRAPPA, Anton Pannekoek Institute for Astronomy, University of Amsterdam,  Science Park 904, 1098 XH Amsterdam, The Netherlands \label{GRAPPA} \and
Friedrich-Alexander-Universit\"at Erlangen-N\"urnberg, Erlangen Centre for Astroparticle Physics, Erwin-Rommel-Str. 1, D 91058 Erlangen, Germany \label{ECAP} \and
Astronomical Observatory, The University of Warsaw, Al. Ujazdowskie 4, 00-478 Warsaw, Poland \label{UWarsaw} \and
School of Physics, University of the Witwatersrand, 1 Jan Smuts Avenue, Braamfontein, Johannesburg, 2050 South Africa \label{WITS} \and
Universit\'e Bordeaux, CNRS/IN2P3, Centre d'\'Etudes Nucl\'eaires de Bordeaux Gradignan, 33175 Gradignan, France \label{CENBG} \and
School of Physical Sciences, University of Adelaide, Adelaide 5005, Australia \label{Adelaide} \and
Universit\"at Hamburg, Institut f\"ur Experimentalphysik, Luruper Chaussee 149, D 22761 Hamburg, Germany \label{HH} \and 
Université de Paris, CNRS, Astroparticule et Cosmologie, F-75013 Paris, France \label{APC} \and
Department of Physics and Astronomy, The University of Leicester, University Road, Leicester, LE1 7RH, United Kingdom \label{Leicester} \and
Nicolaus Copernicus Astronomical Center, Polish Academy of Sciences, ul. Bartycka 18, 00-716 Warsaw, Poland \label{NCAC} \and
Institut f\"ur Physik und Astronomie, Universit\"at Potsdam,  Karl-Liebknecht-Strasse 24/25, D 14476 Potsdam, Germany \label{UP} \and
Laboratoire Univers et Particules de Montpellier, Universit\'e Montpellier, CNRS/IN2P3,  CC 72, Place Eug\`ene Bataillon, F-34095 Montpellier Cedex 5, France \label{LUPM} \and
Landessternwarte, Universit\"at Heidelberg, K\"onigstuhl, D 69117 Heidelberg, Germany \label{LSW} \and
Institut f\"ur Physik, Humboldt-Universit\"at zu Berlin, Newtonstr. 15, D 12489 Berlin, Germany \label{HUB} \and
Institut f\"ur Astro- und Teilchenphysik, Leopold-Franzens-Universit\"at Innsbruck, A-6020 Innsbruck, Austria \label{LFUI} \and
Department of Physics, Rikkyo University, 3-34-1 Nishi-Ikebukuro, Toshima-ku, Tokyo 171-8501, Japan \label{Rikkyo} \and
Institute of Astronomy, Faculty of Physics, Astronomy and Informatics, Nicolaus Copernicus University,  Grudziadzka 5, 87-100 Torun, Poland \label{NCUT} \and
Department of Physics, University of the Free State,  PO Box 339, Bloemfontein 9300, South Africa \label{UFS} \and
Department of Physics, The University of Tokyo, 7-3-1 Hongo, Bunkyo-ku, Tokyo 113-0033, Japan \label{Tokyo} \and
Yerevan Physics Institute, 2 Alikhanian Brothers St., 375036 Yerevan, Armenia \label{YPI} \and
Kavli Institute for the Physics and Mathematics of the Universe (WPI), The University of Tokyo Institutes for Advanced Study (UTIAS), The University of Tokyo, 5-1-5 Kashiwa-no-Ha, Kashiwa, Chiba, 277-8583, Japan \label{KAVLI} 
Now at Physik Institut, Universit\"at Z\"urich, Winterthurerstrasse 190, CH-8057 Z\"urich, Switzerland \label{MitchellNowAt} \and
Now at Institut de Ci\`{e}ncies del Cosmos (ICC UB), Universitat de Barcelona (IEEC-UB), Mart\'{i} Franqu\`es 1, E08028 Barcelona, Spain \label{CerrutiNowAt}
}

\offprints{H.E.S.S.~collaboration, \\
\email{contact.hess@hess-experiment.eu} \\
\protect\\\protect\footnotemark[1] Corresponding authors
\protect\\\protect\footnotemark[2] Deceased
}

\date{Received / Accepted}

\abstract{The unidentified very-high-energy (VHE; E $>$ 0.1 TeV) $\gamma$-ray source, HESS\,J1826$-$130, was discovered with the High Energy Stereoscopic System (HESS) in the Galactic plane. The analysis of 215 h of HESS data has revealed a steady $\gamma$-ray flux from \J1826, which appears extended with a half-width of 0.21$^{\circ}$ $\pm$ 0.02$^{\circ}_{\text{stat}}$ $\pm$ 0.05$^{\circ}_{\text{sys}}$. The source spectrum is best fit with either a power-law function with a spectral index $\Gamma$ = 1.78 $\pm$ 0.10$_{\text{stat}}$ $\pm$ 0.20$_{\text{sys}}$ and an exponential cut-off at 15.2$^{+5.5}_{-3.2}$ TeV, or a broken power-law with $\Gamma_{1}$ = 1.96 $\pm$ 0.06$_{\text{stat}}$ $\pm$ 0.20$_{\text{sys}}$, $\Gamma_{2}$ = 3.59 $\pm$ 0.69$_{\text{stat}}$ $\pm$ 0.20$_{\text{sys}}$ for energies below and above $E_{\rm{br}}$ = 11.2 $\pm$ 2.7 TeV, respectively. The VHE flux from \J1826 is contaminated by the extended emission of the bright, nearby pulsar wind nebula (PWN), HESS\,J1825$-$137, particularly at the low end of the energy spectrum. Leptonic scenarios for the origin of \J1826 VHE emission related to PSR\,J1826$-$1256 are confronted by our spectral and morphological analysis. In a hadronic framework, taking into account the properties of dense gas regions surrounding \J1826, the source spectrum would imply an astrophysical object capable of accelerating the parent particle population up to $\gtrsim$200 TeV. Our results are also discussed in a multiwavelength context, accounting for both the presence of nearby supernova remnants (SNRs), molecular clouds, and counterparts detected in radio, X-rays, and TeV energies.}


\maketitle

\section{Introduction} 

The High Energy Stereoscopic System (HESS) Galactic Plane Survey (HGPS) revealed the presence of 16 new $\gamma$-ray sources in the very high energy (VHE; E $>$ 0.1 TeV) domain \citep{hgps}. Among them, HESS\,J1826$-$130 is located in a region that is exceptionally rich in VHE $\gamma$-ray sources, also encompassing the nearby pulsar wind nebula (PWN) HESS\,J1825$-$137 and the $\gamma$-ray binary system, LS\,5039. HESS\,J1826$-$130 belongs to a growing class of VHE $\gamma$-ray sources affected by source confusion, caused by the influence of luminous neighboring $\gamma$-ray emitters, similar to the case of HESS\,J1641$-$463 \citep{Igor2014}. HESS\,J1826$-$130 is located just $\sim0.7^{\circ}$ to the north of HESS\,J1825$-$137 \citep{hessj1825}, a bright PWN whose extended emission prevented an earlier discovery of HESS\,J1826$-$130 at VHEs.

While HESS\,J1826$-$130 was clearly distinguishable from HESS\,J1825$-$137 in the HGPS, a deep study accounting for its morphological and spectral properties beyond the results obtained with the standard HGPS analysis tools was still lacking. In this paper, the results obtained from the analysis of 215 hours of VHE $\gamma$-ray data recorded with the HESS telescopes is presented, yielding an accurate morphological description and a detailed spectral characterization of the source, including an energy-dependent emission study of the region. The results reported here on HESS\,J1826$-$130 are discussed in a multiwavelength context, discussing in particular its possible association with the spatially coincident supernova remnant, G18.45-0.42, and the Eel Nebula \citep{chandra07}, as well as with the VHE source 2HWC\,J1825$-$134 \citep{2hwccat}, which displays emission up to energies of around 100 TeV. 

Furthermore, HESS\,J1826$-$130 is surrounded by dense molecular clouds that provide support for a hadronic emission scenario. A recent investigation of the diffuse VHE emission around the Galactic center (GC) region demonstrated the presence of PeV particles within the central 10 pc of the Galaxy \citep{gcpev}. This is the first robust detection of a VHE cosmic-ray (CR) accelerator which operates as a PeVatron. The MAGIC Collaboration recently published \citep{magicGC} the results of observations of the diffuse $\gamma$-ray emission in the vicinity of the GC. However, the GC alone would not be able to sustain the CR population close to PeV energies today unless the GC PeVatron was more powerful in the past. This opens up the possibility of other active PeV accelerators in the Galaxy that contribute to the observed Galactic CR flux around the spectral feature known as the \emph{\emph{knee}}, that is, a potentially new CR source population in the Galaxy. HESS\,J1826$-$130 may belong to such a new Galactic CR source population.

In Sect.~\ref{mwl}, we present a summary of the multiwavelength (MWL) information available in the field of view (FoV) around HESS\,J1826$-$130. The HESS observations and data analysis are reported in Sect.~\ref{hess}, including morphological and spectral results, as well as a study of the contamination from the nearby PWN HESS\,J1825$-$137. Results are discussed in Sect.~\ref{interpret}, with concluding remarks in Sect.~\ref{conc}.

\section{Multiwavelength environment}\label{mwl}

\begin{figure*}[!ht]
\centering
\includegraphics[width=12cm]{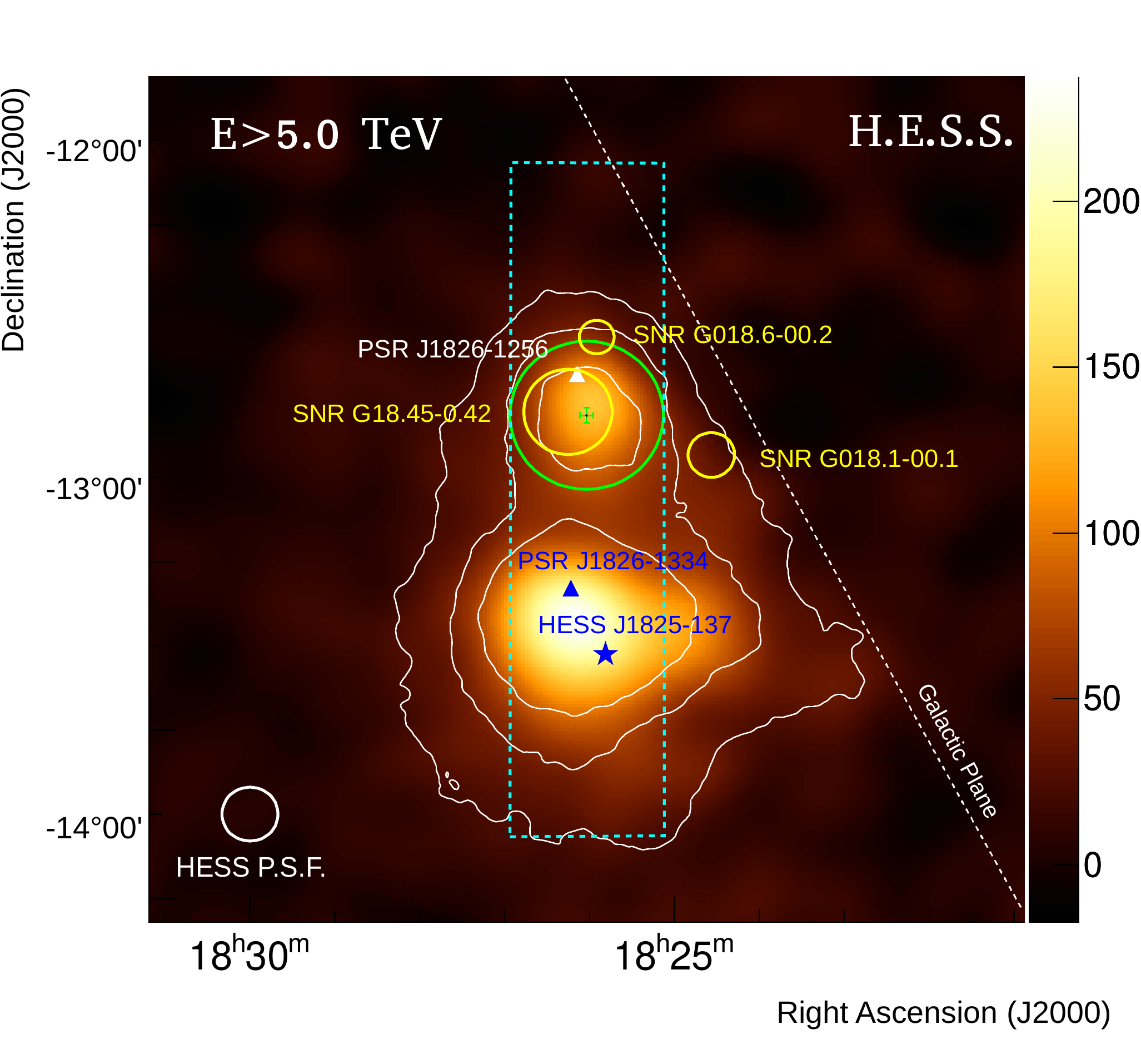}
\caption{Map of excess counts with energies E $>$ 5 TeV for region around HESS\,J1826$-$130 smoothed with a Gaussian kernel of 0.07$^{\circ}$ to match HESS PSF at the same energies. The 68$\%$ containment radius of the HESS PSF is shown with the white circle in the lower-left corner. The white contours indicate the significance of the emission at the 5$\sigma$, 10$\sigma,$ and 15$\sigma$ level. The color scale is in units of excess counts per smoothing Gaussian width. The green circle shows the integration region used for deriving the source spectrum (Fig. \ref{spectrum}), while the green cross indicates the value and 1$\sigma$ uncertainty of the best-fit position of the source. The nearby SNRs: G18.45$-$0.42, G018.6$-$00.2, and G018.1$-$00.1 are represented with yellow circles, while the white triangle indicates the position of the $\gamma$-ray pulsar PSR J1826$-$1256. The blue star shows the peak position of the bright nearby source HESS\,J1825$-$137 for energies above 0.25 TeV (we employed the position as reported in \cite{hessj1825} since the data set analyzed here accounts for a total exposure across all of HESS\,J1825$-$137, which is less than that used in its dedicated study). The blue triangle indicates the position of PSR J1826$-$1334. The white dashed line indicates the orientation and position of the Galactic plane, while the cyan dashed rectangle indicates the area for the extraction of the profiles shown in Fig.~\ref{slices}. Color version of Fig.~\ref{excess} available in the electronic version. }
\label{excess}
\end{figure*}

As mentioned before, HESS\,J1826$-$130 is located in a region with several VHE $\gamma$-ray sources. At VHEs, the TeV sky around the source is dominated by one of the brightest and largest PWNs detected at VHEs, HESS\,J1825$-$137. This latter source is thought to be powered by PSR J1826$-$1334 (also known as PSR\,B1823$-$13), a powerful pulsar with a spin-down luminosity of $\sim3 \times 10^{36}$~erg~s$^{-1} $, located about 13' from the peak position of HESS\,J1825$-$137 in the VHE band. At GeV energies, both the spectral and the morphological properties of HESS\,J1825$-$137 resemble and smoothly connect with those observed at VHEs, as reported in \citet{Grondin2011} using the $Fermi$-LAT.

The first evidence for $\gamma$-ray emission originating from the north of PSR\,J1826$-$1334 was found by the detection of 3EG\,J1826$-$1302 (also called GeV\,J1825$-$1310) in 1999 with the EGRET telescope on-board the Compton Gamma Ray Observatory \citep{egretcat}. Follow-up X-ray observations of the same region led to the detection of a diffuse X-ray source, AX\,J1826.1$-$1300, in the 2$-$10\, keV energy band \citep{ascacat}. This X-ray source is characterized by two emission peaks located on top of a diffuse X-ray nebula. Further observations of AXJ1826.1$-$1300 in 2007 with the Chandra X-ray Observatory resolved the source in more detail (\citet{chandra07}). The southern peak was found to be likely associated with a stellar cluster, whereas the northern peak was resolved and associated with a PWN. Due to a $4^{\prime}$ trail of hard X-ray emission connected to this new nebula, the authors called it the Eel Nebula. \citet{chandra07} were the first to introduce a new separate VHE $\gamma$-ray source north of HESS\,J1825$-$137, which they named HESS\,J1826$-$131 and associated with the Eel Nebula. Until that time, the remaining VHE emission to the north of HESS\,J1825$-$137 was thought to be connected to the bright source itself. 

The \emph{Fermi}/LAT detected the radio-quiet $\gamma$-ray pulsar PSR\,J1826$-$1256 \citep{fermipsr09}, with a spin down luminosity $\dot{\text{E}}$ = 3.6 $\times$ 10$^{36}$ erg s$^{-1}$ and a characteristic age $\tau_{\text{c}}$\,=\,14.4\,kyr, whose position is consistent with the Eel Nebula, representing a significant step forward in our understanding of the various sources contributing to the $\gamma$-ray emission from this region. In 2017, HESS\,J1826$-$130 was reported as a new source within the HGPS. The HGPS used a dedicated source-identification algorithm based on a multi-Gaussian component fit to a given region of interest covered by the survey \citep{hgps}.
In particular for HESS\,J1826$-$130, the spatially extended VHE $\gamma$-ray emission region matched the position of 3EG\,J1826$-$1302, also overlapping with the diffuse X-ray nebula observed with ASCA (which displays a radius of about $15^{\prime}$; see Fig.~11 in \citep{Roberts2001}) as well as the Eel Nebula (with an extension of about $4^{\prime}$ \citep{chandra07}).
The latest results at VHEs of the region include the significant detection of an emission region, 2HWC\,J1825$-$134, included in the 2$^{\rm nd}$ HAWC catalog \citep{2hwccat}. Due to the large spatial extension of the HAWC emission region and the relatively low angular separation between HESS\,J1826$-$130 and HESS\,J1825$-$137, the two sources were not distinguishable in this catalog. Further studies, however, using a much larger dataset and employing updated analysis methods, provide a statistical improvement at the level of $\sim 5 \sigma$ when accounting for the presence of the two HESS VHE sources.

Within the surroundings of HESS\,J1826$-$130, there are also three supernova remnants (SNRs): G18.1$-$0.1, G18.6$-$0.2, and G18.45$-$0.42, which are detected in the radio band \citep{brogan06, Anderson2017}. Detailed studies of the gas dynamics and densities in the direction of the region north of HESS\,J1825$-$137 were carried out by \citet{voisin16}, following the discovery of a molecular cloud north of PSR\,J1826$-$1334 by \citet{lemiere06}. These studies revealed the presence of dense molecular clouds with particle density values up to $n_{\mathrm{H}_2}$\,\textasciitilde\,$7\times10^2\,\mathrm{cm}^{-3}$. The bulk of molecular gas in the vicinity of HESS\,J1826$-$130 is located at V$_\mathrm{LSR}$ = 45--60\,km/s, corresponding to a kinematic distance of 4.0 kpc, and V$_\mathrm{LSR}$ = 60--80\,km/s, corresponding to a kinematic distance of 4.6 kpc. The former gas distance estimate is consistent with the dispersion measurement of PSR\,J1826$-$1334. \citet{voisin16} concluded that due to the general spatial match between the molecular gas and the TeV emission, HESS\,J1826$-$130 might be explained by a hadronic emission scenario, with the progenitor SNR of HESS\,J1825$-$137 being a suitable local CR source candidate. However, a leptonic scenario connected to the Eel Nebula could not be ruled out. Furthermore, the two SNRs, G18.1$-$0.1 and G18.6$-$0.2, were considered unlikely to be directly related to the TeV excess due to their offset positions and their small angular diameters.

\section{HESS observations and data analysis results} \label{hess}

The High Energy Stereoscopic System is an array of five imaging atmospheric Cherenkov telescopes located in the Khomas Highland of Namibia, 1800 m above sea level. HESS in phase I comprised four 12 m diameter telescopes that have been fully operational since 2004. A fifth telescope (CT5), with a larger mirror diameter of 28 m and newly designed camera \citep{hess2}, was added to the center of the array and has been operational since September 2012. The HESS phase\,I array configuration is sensitive to $\gamma$-ray energies between 100\,GeV and several tens of TeV, while the energy threshold of the array was lowered to 10's of GeV with the addition of CT5. The VHE HESS data presented in this paper were recorded with the HESS phase\,I array configuration, which can measure extensive air showers with an angular resolution better than ${0.1}^{\circ}$ and an average energy resolution of 15$\%$ for an energy of 1 TeV \citep{Aharonian06}.

The VHE $\gamma$-ray observations of the FoV around HESS\,J1826$-$130 were carried out between 2004 and 2015. About 140 h of observations were recorded with the HESS phase I array configuration (between 2004 and 2012), while $\sim$95\,h were obtained with the HESS phase II array configuration in a split observation mode without participation of the CT5 telescope. These data sets provide a total acceptance-corrected live-time of 215 h of HESS data after the application of quality selection criteria \citep{Aharonian06}.

The data have been analyzed with the HESS Analysis Package (HAP) for shower reconstruction, and a multivariate analysis technique \citep{TMVA} was applied to provide an improved distinction between hadron and $\gamma$-ray events. A cross-check analysis was performed using an independent calibration and analysis method \citep{m++}, which gives compatible results with the main analysis. 

\subsection{Detection and morphological analysis} \label{morph}

In order to achieve an improved angular resolution and reduce strong contamination coming from the nearby bright source, HESS\,J1825$-$137, the source position and morphology of HESS\,J1826$-$130 were obtained with a cut configuration optimized for high energies that requires a minimum of 160 photo-electrons per image. The cosmic-ray background level was estimated using the ring background model \citep{berge2007}. Using the full 215 h dataset, HESS\,J1826$-$130 was detected with a statistical significance of 22.6$\sigma$\footnote{For comparison, HESS\,J1826$-$130 is detected at a statistical significance of about 9.4~$\sigma$ in the HGPS, \citep{hgps}.}, determined following Equation (17) in \cite{lima}. Figure \ref{excess} shows the acceptance-corrected VHE $\gamma$-ray excess map of the region around HESS\,J1826$-$130 at energies greater than 5 TeV, smoothed with the HESS point spread function (PSF) with a radius of 0.07$^{\circ}$. 

\begin{figure}[t!]
\resizebox{\hsize}{!}{\includegraphics[width=0.55 \textwidth]{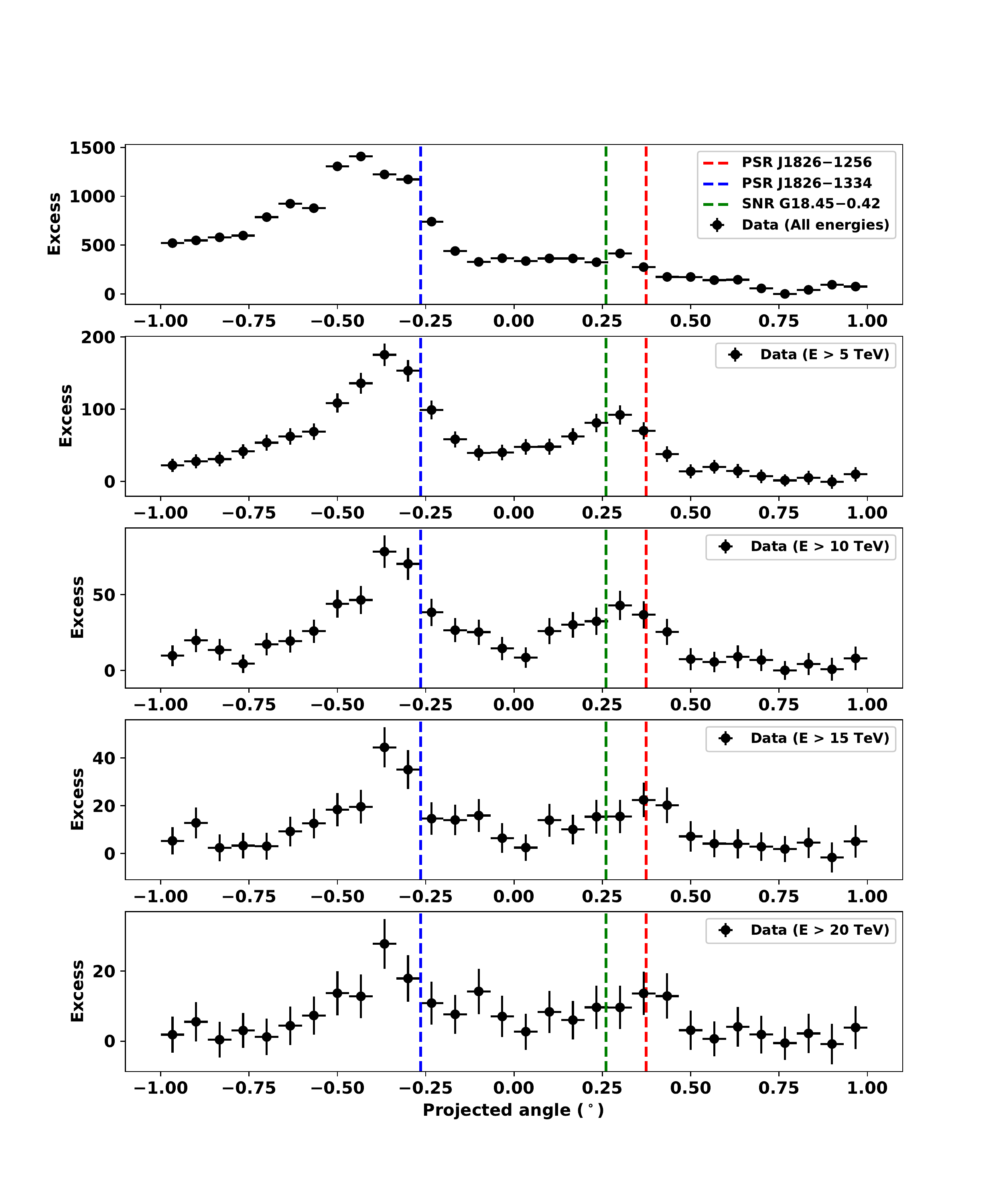}}
\caption{Distribution of VHE $\gamma$-ray excess along axis of the cyan dashed rectangle indicated in Fig.~\ref{excess} that encompasses both HESS\,J1825$-$137 (broad peak on the left) and  HESS\,J1826$-$130 (shallower peak on the right, becoming apparent only at high energies) above 0.4 TeV and at $E> 5$ TeV, 10 TeV, 15 TeV, and 20 TeV. Vertical lines show the position of PSR\,J1826$-$1256 (red), PSR\,J1826$-$1334 (blue) and SNR\,G018.45$-$0.42 (red).}
\label{slices}
\end{figure}

Until 2017, in the first two publications by the HESS Collaboration concerning HESS\,J1825$-$137 (\citet{oldHessGPS}; \citet{oldJ1825Paper}), HESS\,J1826$-$130 remained unnoticed and hidden to the standard source detection techniques due to its low brightness with respect to HESS\,J1825$-$137. Confirmed as a separate source in 2017 \citep{hgps}, HESS\,J1826$-$130 emerges toward higher energies as illustrated in Fig.~\ref{slices}, with reduced contamination from HESS\,J1825$-$137, potentially due to its energy-dependent morphology and hinting at a harder VHE spectrum.

The best-fit position and extension of the source have been determined by convolving 2D Gaussian models with the HESS PSF and fitting this to the excess event distribution (for E $>$ 0.4 TeV). The HESS PSF was derived from simulations and modeled as a weighted sum of three 2D Gaussian functions for the morphology analysis. A log-likelihood ratio test (LLRT) was used to select the best morphology model that represents the HESS data. The centroid of the 2D Gaussian corresponding to the best-fit position of HESS\,J1826$-$130 is R.A. (J2000): $18^{\textsuperscript{h}}$26$^{\textsuperscript{m}}$02.16$^{\textsuperscript{s}}$ $\pm$ 3.38$^{\textsuperscript{s}}_{\rm stat}$ $\pm$ 1.30$^{\textsuperscript{s}}_{\rm sys}$ and Dec. (J2000): $-$13$^{\circ}$04$\arcmin$ $\pm$ 1.0$\arcmin_{\rm stat}$ $\pm$ 0.3$\arcmin_{\rm sys}$. The extension (radius) of the source is estimated to be 0.21$^\circ$ $\pm$ 0.02$^{\circ}_{\rm stat}$ $\pm$ 0.05$^{\circ}_{\rm sys}$, which is significantly larger than the HESS PSF. 

\begin{table} [t] 
\renewcommand{\arraystretch}{1.3}
\centering
\tiny
\caption{Analysis statistics of the HESS\,J1826$-$130 region using different energy thresholds ($E_{\rm th}$) corresponding to the VHE $\gamma$-ray excess profiles shown in Fig.~\ref{slices}. The events below the energy threshold shown in the $E_{\rm th}$ column were not taken into account in the analysis. The number of events within the source region (N$_\text{On}$), within the ring region used for estimating the background level (N$_\text{Off}$), the alpha factor ($\alpha$), the corresponding number of excess events, and the Li$\&$Ma statistical significance \citep{lima} are given for each energy threshold separately.}
\begin{tabular}{c c c c c c c}
\hline 
\hline 
$E_{\rm th}$ (TeV)  & N$_\text{On}$ & N$_\text{Off}$ & $\alpha$ & Excess & Significance ($\sigma$)  \\ 
\hline 

0.4           & 5544          & 27006    & 0.145    & 1627.6   & 22.7   \\
5             & 569           & 1685     & 0.132    & 346.2    & 17.8   \\
10            & 237           & 664      & 0.126    & 153.4    & 12.6   \\ 
15            & 118           & 363      & 0.125    & 72.5     & 8.2    \\ 
20            & 68            & 220      & 0.120    & 41.6     & 6.2    \\ 

\hline
\hline
\end{tabular}
\label{engStatTable}
\end{table}

\subsection{Spectral analysis} \label{spect}

\begin{table*}[ht!]
  \begin{center}
    \renewcommand{\arraystretch}{1.3}
    \caption{Comparison of different spectral models fit to the data in the energy range of [0.4, 56.2] TeV, together with the spectral parameter values. In all cases, statistical errors are quoted, while systematics errors are 20$\%$ for flux and 0.2 for spectral indices. The likelihoods and p-values are given for each model for comparison.}
    \label{spect_table}
    \begin{tabular}{c c c c c c c} 
      \hline
      \hline

     Spectral model & Flux (1 TeV) & Int. flux($>$ 1 TeV) & Parameters & -2*Loglike & $\chi^{2}/n.d.f.$ \\
                    & $\times$$10^{-13}$ cm$^{-2}$s$^{-1}$TeV$^{-1}$ & $\times$$10^{-13}$ cm$^{-2}$s$^{-1}$ & & (LLRT) & (p-val) \\

      \hline
      Power law    & 11.1 $\pm$ 0.6 & 9.8 $\pm$ 0.4   & $\Gamma$ = 2.12 $\pm$ 0.04     & 44.015         & 37.3 / 10  \\
                  &                & (4.7$\%$ Crab)  &                                & (nested)       & (5$\times$$10^{-5}$) \\

      \hline
      Exp. cut-off & 10.1 $\pm$ 0.7 & 10.2 $\pm$ 0.5 & $\Gamma$ = 1.78 $\pm$ 0.10      & 33.566         & 9.3 / 9 \\
      power law    &                & (4.9$\%$ Crab) & $\lambda$ = 0.0659 $\pm$ 0.0177 (TeV$^{-1}$) & (3.2 $\sigma$) & (0.41)  \\
      \hline
      Broken      & 10.2 $\pm$ 0.7 & 10.1 $\pm$ 0.5 & $\Gamma_{1}$ = 1.96 $\pm$ 0.06  & 32.577         & 8.6 / 8 \\
      power law    &                & (4.8$\%$ Crab) & $\Gamma_{2}$ = 3.59 $\pm$ 0.69  &  (n/a)         & (0.38)  \\
                  &                &                & E$_{\rm br}$ = 11.2 $\pm$ 2.7 (TeV)      &                &         \\

      \hline
      \hline

    \end{tabular}
  \end{center}
\end{table*}

\begin{figure}[t]
\resizebox{\hsize}{!}{\includegraphics[width=0.55 \textwidth]{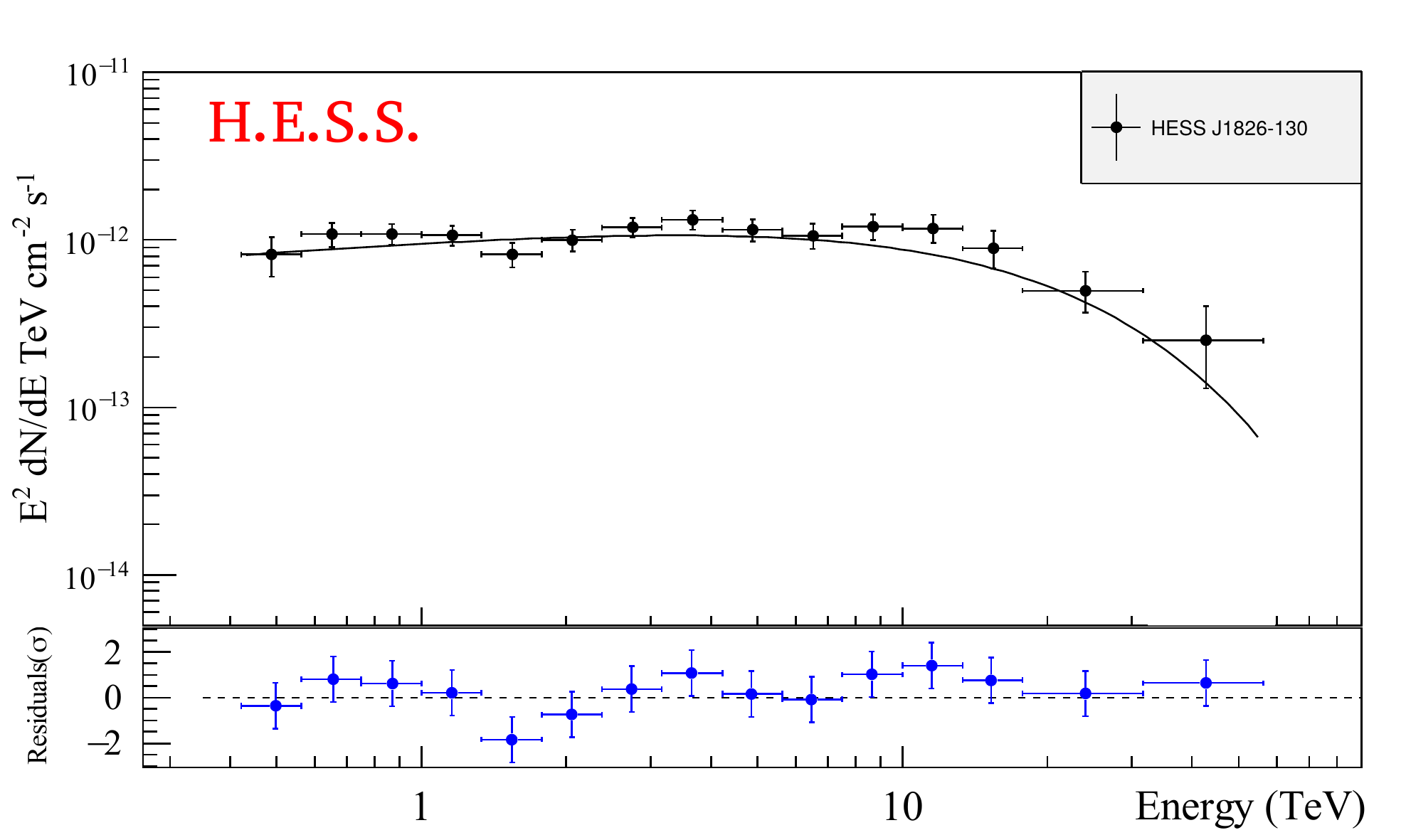}}
\caption{VHE $\gamma$-ray spectrum of HESS\,J1826$-$130 extracted from the green circular source region shown in Fig. \ref{excess}. The black dots show the flux points with 1$\sigma$ statistical errors. The 1$\sigma$ and 3$\sigma$ error bands of the best-fit ECPL model are shown with the blue and light blue shaded regions, respectively, while the best-fit ECPL model itself is shown with the black line. The spectrum shown is binned such that all flux points have a minimum significance level of 2.0$\sigma$. The significance of the last flux bin centered at 42.7 TeV is 2.4$\sigma$.}
\label{spectrum}
\end{figure}

The green circular region with a radius of 0.22$^{\circ}$ centered at the best-fit position of HESS J1826$-$130 (see Fig. \ref{excess}) was used as the integration region for measuring the differential VHE $\gamma$-ray spectrum of the source. The reflected background model \citep{berge2007} was used to estimate the background. The source spectrum was derived using the forward folding technique \citep{piron2001}. 
Several spectral models were used to fit the data, including a simple power-law (PL) model, a broken power-law (BPL) model, and a power-law model with an exponential cut-off (ECPL; see Table \ref{spect_table}). The spectral fit is performed between 0.42 TeV and 56.2 TeV. The best-fit model is found to be the ECPL, with a $p$-value = 0.41 and a likelihood ratio test (LLRT) against the PL model providing a statistical improvement at the $\sim 3.2 \sigma$ level. The differential VHE emission is thus best represented by $dN/dE$ = $\Phi_{0}$ ($E/1\textnormal{ TeV}$)$^{-\Gamma}$ $\exp$($-E/E_{\text{c}}$), with $\Phi_{0}$ = (10.10 $\pm$ 0.69$_{\rm stat}$ $\pm$ 2.02$_{\rm sys}$) $\times$ $10^{-13}$ cm$^{-2}$ 
s$^{-1}$ TeV$^{-1}$, $\Gamma$ = 1.78 $\pm$ 0.10$_{\rm stat}$ $\pm$ 0.20$_{\rm sys}$ and a cut-off energy of $E_{c}$ = 15.2$^{+5.5}_{-3.2}$ TeV as shown in Fig. \ref{spectrum}. The integral flux level above 1 TeV is $\Phi$($>$ 1 TeV) = (10.2 $\pm$ 0.5$_{\rm stat}$ $\pm$ 2.0$_{\rm sys}$) $\times$ $10^{-13}$ cm$^{-2}$ s$^{-1}$ and corresponds to 4.9$\%$ of the Crab Nebula flux at the same energies\footnote{Throughout this paper, 1 Crab unit is defined here as $\Phi$($>$ 1 TeV) = 2.26 $\times$ $10^{-11}$ cm$^{-2}$ s$^{-1}$ \citep{Aharonian06}.}. A 2$\sigma$ lower limit on the cut-off energy of the source is derived at $\sim10$ TeV. A fit to a BPL model gives a break energy at 11.2 $\pm$ 2.7 TeV, which is compatible with the cut-off energy obtained from the ECPL model. The comparison between the ECPL and the BPL model does not show any preference from obtained p-values. Both models clearly indicate a spectral steepening of the spectrum above 10 TeV. The integral flux is found to be constant within the HESS dataset at different timescales, from run-wise ($\sim$30 min) to year-wise light curves. 

The integration region used to measure the VHE spectrum of HESS\,J1826$-$130 is strongly contaminated, especially at lower energies, by the relatively softer spectrum of HESS\,J1825$-$137. Therefore, the obtained spectrum is affected by this contamination. Taking into account the relatively softer contribution from the contaminating source, the intrinsic spectrum of HESS\,J1826$-$130 is expected to be even harder with respect to the spectral results discussed above.  

\subsection{Contamination study} \label{cont}

The estimation of intrinsic spectra for sources strongly contaminated by neighboring bright and extended sources presents a nontrivial challenge. Although morphology studies in different energy bands can provide a powerful tool for new discoveries, such a method does not help in providing a quantitative characterization of the contamination, as systematic uncertainties in the flux estimation are expected to dominate. In general, contamination effects can be neglected if they are at the $10\%$ level or lower, since this is, for example, the typical HESS systematic error on the source flux. For contamination exceeding $10\%$, in particular for spectral studies, a dedicated analysis is needed. 

\begin{table*}[!ht] 
\renewcommand{\arraystretch}{1.3}
\caption{Comparison of ECPL spectral models for the HESS\,J1826$-$130 integration region (Total), plus the north and south semi-circle regions. The integral flux (in $\%$ Crab Nebula) is given for the same energy range (above 1 TeV).} 
\label{morphtable}
\centering
\begin{tabular}{c c c c c c} 
\hline
\hline 
Region             & Normalization                                        & Index                                               & Cut-off energy        & Integral flux ($>$1 TeV)  \\ 
                   & ($10^{-13}$ cm$^{-2}$ s$^{-1}$ TeV$^{-1}$)           &                                                     & (TeV)                 & ($\%$ Crab Nebula)          \\
\hline 
Total              & 10.10 $\pm$ 0.69$_{\rm stat}$ $\pm$ 2.02$_{\rm sys}$ & 1.78 $\pm$ 0.10$_{\rm stat}$ $\pm$ 0.20$_{\rm sys}$ & 15.2$^{+5.5}_{-3.2}$  & 4.9 \\ 
North semi-circle  & 4.64 $\pm$ 0.49$_{\rm stat}$ $\pm$ 0.93$_{\rm sys}$  & 1.69 $\pm$ 0.14$_{\rm stat}$ $\pm$ 0.20$_{\rm sys}$ & 15.8$^{+9.6}_{-4.3}$  & 2.4 \\ 
South semi-circle  & 5.06 $\pm$ 0.49$_{\rm stat}$ $\pm$ 1.01$_{\rm sys}$  & 1.88 $\pm$ 0.14$_{\rm stat}$ $\pm$ 0.20$_{\rm sys}$ & 16.4$^{+11.9}_{-4.7}$ & 2.3 \\ 
\hline
\hline 
\end{tabular}
\label{spectCompTab}
\end{table*}

Performing contamination studies for HESS\,J1826$-$130 requires knowledge about the energy-dependent morphology of HESS\,J1825$-$137, which is highly asymmetrical. The contamination was estimated in this case using the morphology model that represents the HESS data best fit. The model includes a 2D Gaussian function accounting for HESS\,J1826$-$130 along with other 2D Gaussian functions representing the other sources, HESS\,J1825$-$137 and the gamma-ray binary system LS 5039, present in the FoV. Taking the ratio of the total number of excess counts within the integration region (Fig. \ref{excess}, green circle) used for measuring \J1826's spectrum between the model components can give a quantitative estimation of the contamination. By using such an approach, the contamination from HESS\,J1825$-$137 is estimated to be $\sim$40$\%$ below 1.5 TeV and $\sim$20$\%$ above 1.5 TeV, while it is strongly reduced (<$10\%$) at energies above 2.0 TeV. Given the relatively large angular distance of LS 5039 to the integration region (around 2.0$^{\circ}$) encompassing HESS\,J1826$-$130, we find that the contamination from the binary system is negligible.

\begin{figure}[t]
\begin{center}
\includegraphics[width=0.45\textwidth]{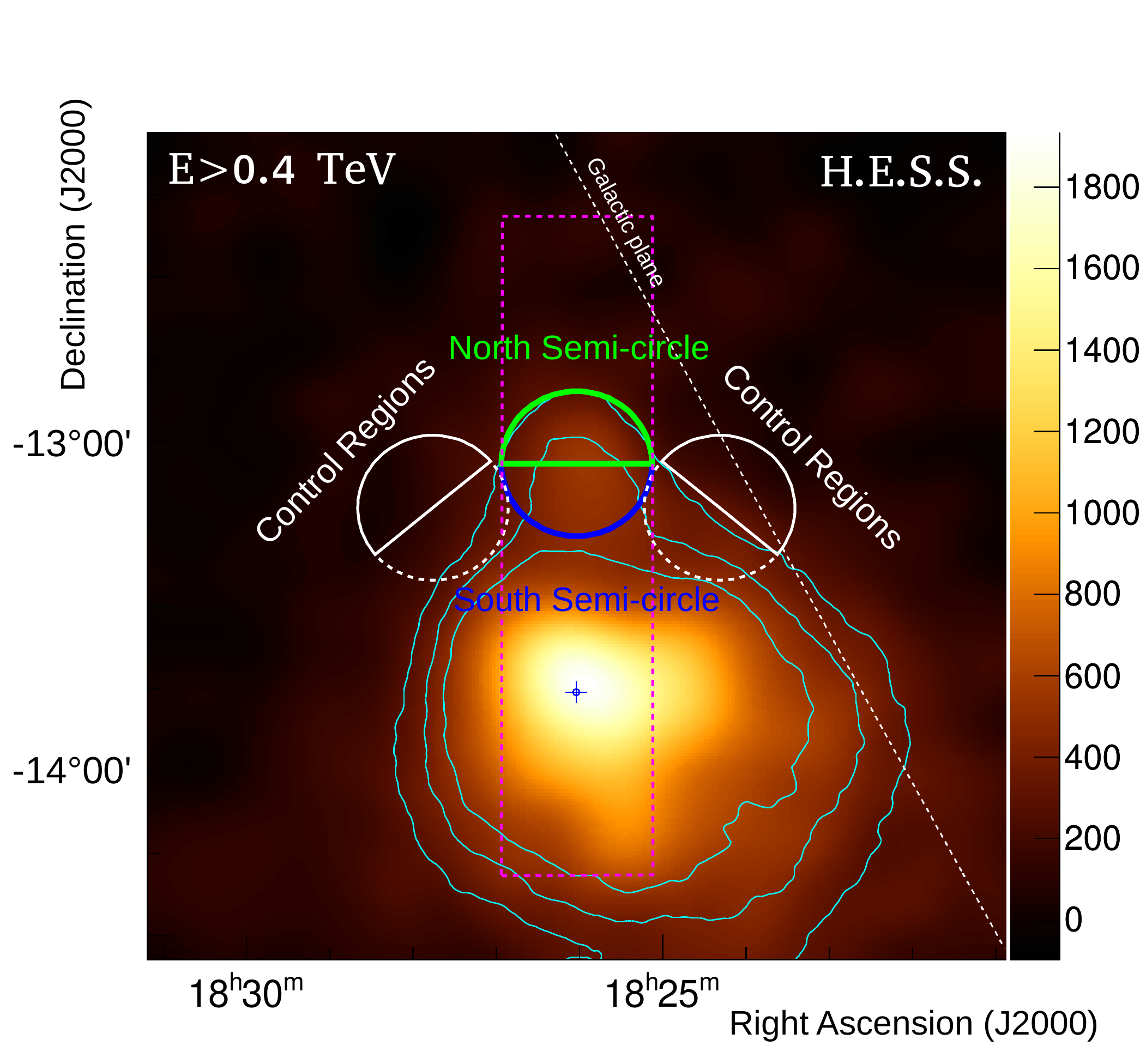}
\caption{VHE $\gamma$-ray excess map of HESS\,J1826$-$130 for all events saturated for a better visualization, showing the north semi-circle (green), south semi-circle (blue), and control regions for both north (solid white) and south (dashed white) semi-circle regions. The control regions are distributed in a radially symmetric manner around the peak position of HESS\,J1825$-$137, indicated with the blue cross, obtained from the dataset and for energies above 0.4 TeV. The cyan contours indicate the significance of the emission at the 15$\sigma$, 20$\sigma$ and 25$\sigma$ levels.}
\label{NS_Cont}
\end{center}
\end{figure}

The spectral contamination from HESS\,J1825$-$137 has been estimated based on the fact that as the separation between contaminated and contaminating source increases, one expects the effect of contamination to decrease. To test the biasing impact of the contamination, the HESS\,J1826$-$130 integration region was split into a near and far half-circle, referred to as the south and north semi-circles shown in Fig. \ref{NS_Cont}. Spectra from two control regions at both sides of the source position were also extracted in order to determine whether or not the contribution from HESS\,J1825$-$137 for the north (south) semi-circle is significant. The spectra of the upper halves of these control regions (solid semi-circles in Fig. \ref{NS_Cont}) show insignificant (< 5.0$\sigma$) emission, while the lower halves of the control regions (dashed semi-circles in Fig. \ref{NS_Cont}), which are closer to HESS\,J1825$-$137, show significant (> 5.0$\sigma$) VHE $\gamma$-ray emission. 

One expects that the contamination in the north semi-circle region should be less with respect to the south semi-circle region, and the spectrum should reflect the intrinsic spectral properties of HESS\,J1826$-$130. The spectral fit results, assuming an ECPL model and obtained from the north and south semi-circle regions, are given in Table \ref{spectCompTab} along with the total results. As expected, the north semi-circle spectrum is slightly harder than those of the total and south semi-circle regions. Although the contamination from HESS\,J1825$-$137 could potentially affect the spectrum of HESS\,J1826$-$130, our studies demonstrate that the effects on the spectral parameters are at the level of our analysis systematic uncertainties. The intrinsic spectrum of HESS\,J1826$-$130, corrected for this contamination, is found to be well described by an ECPL model where $\Phi_{0}$ = 9.2 $\times$ $10^{-13}$ cm$^{-2}$ s$^{-1}$ TeV$^{-1}$, $\Gamma$ = 1.7 and a cut-off energy around $E_{c}$ = 16 TeV. 

A further investigation of the intrinsic spectrum of HESS\,J1826$-$130 was performed using a 3D cube-analysis approach\footnote{We used ctools \citep{ctools} and gammapy \citep{gammapy}, which include 3D cube analysis approaches optimized for measuring intrinsic spectra from contaminated sources.}. With such a template-based analysis method, uncontaminated spectra can be derived, since all sources located in the FoV are simultaneously taken into account. The 3D analysis results published in \cite{alexPhd} are found to be compatible with the results presented here.

\section{Discussion} \label{interpret}

\begin{figure}[t]
\begin{center}
\includegraphics[width=0.45\textwidth]{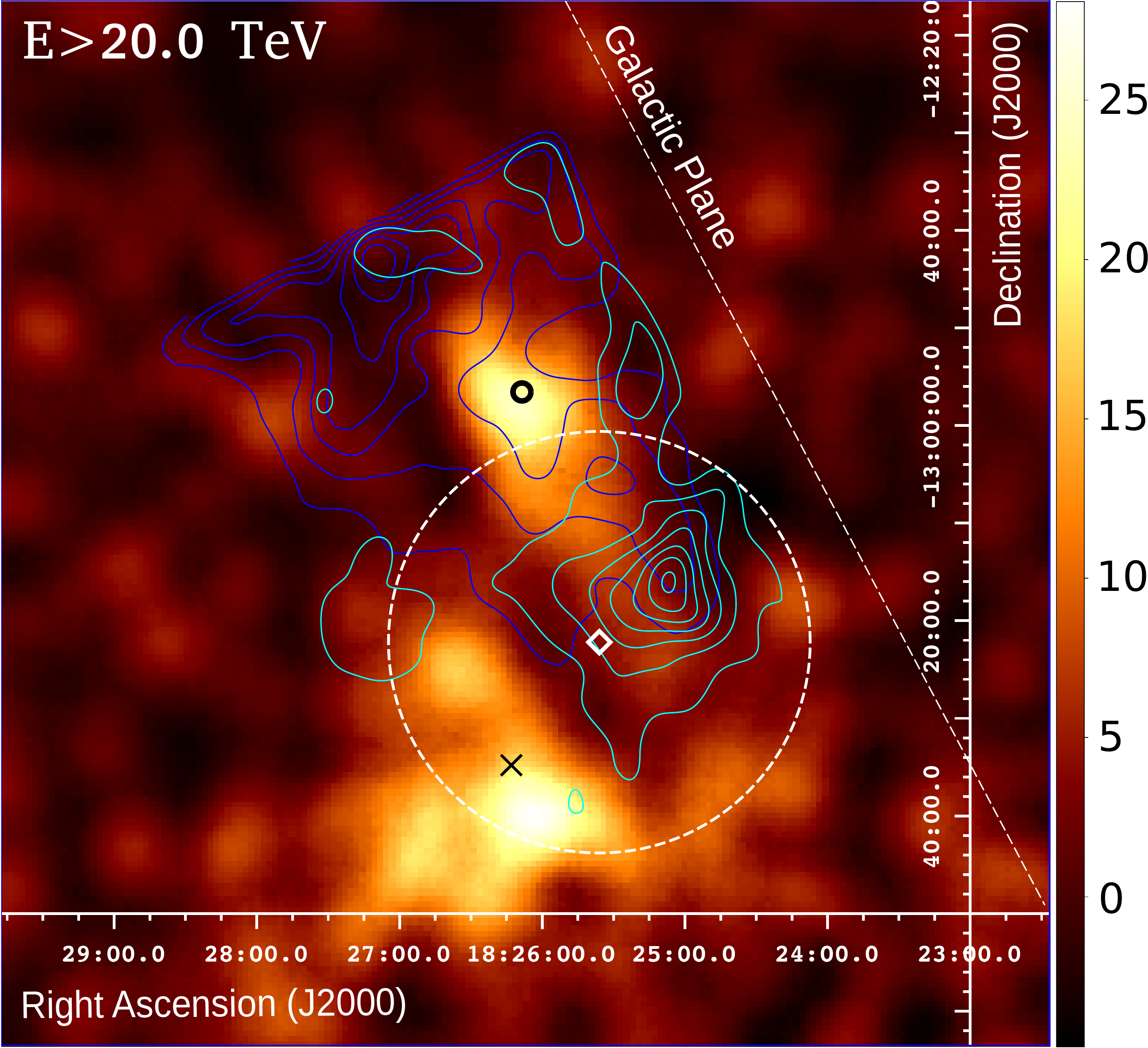}
\caption{VHE $\gamma$-ray excess map of HESS\,J1826$-$130 for events above 20 TeV, together with total column density map contours. The black circle and cross show the location of the pulsars PSR\,J1826$-$1256 and PSR\,J1826$-$1334. The cyan and blue contours show the total column density contours of [2.5, 9.5] $\times$ 10$^{22}$ cm$^{2}$, derived from Nobeyama $^{12}$CO(1--0) and SGPS HI data, for gas velocities between 45--60 km/s (d: 4 kpc) and 60--80 km/s (d: 4.6 kpc), respectively. The white dashed circle and diamond are the extension and best-fit position of recently reported $E > 100$~TeV emitting source, eHWC J1825$-$134 \citep{HAWC2019}.}
\label{E_20TeV_skymap}
\end{center}
\end{figure}

The study of VHE $\gamma$-ray sources displaying spectra extending above several tens of TeV is required to understand the origin of the highest energy CRs close to the knee in the CR spectrum. These studies are particularly promising in cases in which dense gas regions are coincident with the source of interest. In the case of HESS\,J1826$-$130, the spatial coincidence of a dense molecular hydrogen region found along the line of sight (see \citet{voisin16}), suggests that the radiation could actually be produced by protons with energies of several hundred TeV colliding with this gas. 
Figure~\ref{E_20TeV_skymap} displays the HESS excess map obtained at energies above 20 TeV together with column density profiles derived from Nobeyama $^{12}$CO(1--0) \citep{nobeyama} and SGPS HI data \citep{sgps}, for two different gas velocity ranges, 45--60 km/s (corresponding to a kinematic distance of 4.0 kpc) and 60--80 km/s (corresponding to a kinematic distance of 4.6 kpc). The scaling X-factors, X$_{\textnormal{HI}}$ = 1.8 $\times$ $10^{18}$ cm$^{-2}$ (K km s$^{-1}$)$^{-1}$ and X$_{\textnormal{CO}}$ = 2.0 $\times$ $10^{20}$ cm$^{-2}$ (K km s$^{-1}$)$^{-1}$, were taken from \citet{hi_factor} and \citet{co_factor}, respectively. The Nobeyama $^{12}$CO(1--0) data was smoothed to match the SGPS beam size of 2'. The CO and HI distribution display two local maxima that do not coincide with the $E>20$~TeV peak, although they still provide target densities at the level of $\sim 5 \times 10^{2}$~cm$^{-3}$ at the position of HESS\,J1826$-$130.

The best-fit parent proton population in this scenario would have a spectral index of $\sim$1.7 and a cut-off energy around 200 TeV\footnote{We employed the NAIMA fitting software package for these computations \citep{Zabalza2015}.}.
The relative hardness of the photon spectrum could be due to the highest energy protons diffusing into dense clouds, while lower energy protons might still be confined within the accelerating source, or efficiently excluded from the clouds if the diffusion coefficient inside the clouds is suppressed \citep{sabrina}. 
A hadronic scenario in which runaway protons accelerated by one of the coincident or nearby SNRs emit TeV photons when interacting with the dense ambient gas found along the line of sight could explain the hard spectrum of HESS\,J1826$-$130. We note, however, that the filled centers of SNR G18.5-0.04 and SNR G18.1-00.1 are possibly too old to be able to accelerate protons up to hundreds of TeV.
The $\gamma$-ray luminosity, L$_{\gamma}$, of the source is 8 $\times$ 10$^{33}$ erg s$^{-1}$ for a distance of 4 kpc, corresponding to the estimated distances of the SNRs G18.45$-$0.42 \citep{Karpova2019}, G018.1$-$00.1, or G018.6$-$00.2 \footnote{We note that \cite{voisin16} discuss the possibility of a larger distance to G018.6$-$00.2 of 11.4 kpc, disfavoring an association of HESS\,J1826$-$130 with G018.6$-$00.2.}. This would translate into an energy output in accelerated protons W$_{\textnormal{pp}}$ = L$_{\gamma}$ $\times$ t$_{\textnormal{pp}}$, of 6 $\times$ 10$^{49}$ ($n$/1 cm$^{-3}$)$^{-1}$ erg, where the gas density value is $n$ = $5\times10^2\,\mathrm{cm}^{-3}$ and t$_{\textnormal{pp}}$ is the typical timescale for gamma-rays to be produced through neutral pion decay following the collision of relativistic protons. 

The relatively hard spectrum found at VHE gamma rays could also be produced in a leptonic scenario by an electron population with a spectral index close to 2.0 and a cut-off at around 70 TeV. These electrons could be accelerated by the pulsar PSR\,J1826--1256 and up-scatter cosmic microwave background (CMB) and IR photons to produce the observed TeV emission. PSR\,J1826$-$1256 is powering the Eel Nebula (PWN G018.5--0.4, \citealp{chandra07}), which displays an elongated X-ray trail of shocked material associated with its fast proper motion through the surrounding medium. An estimate for the magnetic field inside the Eel Nebula of $\sim$20--30 $\mu$G is provided in \cite{Keogh2010}. 
Recently, \cite{Duvidovich2019} made use of XMM-$Newton$ observations of the region around PSR\,J1826--1256 and discussed such a possibility. Based on the spectral softening found along the nebula, they considered X-rays produced by electrons with energies $E_{\rm e}\lesssim 150$~TeV radiating synchrotron emission in a magnetic field $B > 2 \mu$G. In a parallel study, \cite{Karpova2019} also reported on the X-ray analysis of this very same XMM-$Newton$ dataset, complemented with additional $\sim90$~ks of $Chandra$ observations of the region. These data support the association of HESS\,J1826$-$130 with the Eel Nebula, although somewhat harder spectral indices close to the PWN are derived ($\Gamma_{\rm X}\sim1.2)$, followed by softer emission, with $\Gamma_{\rm X}\sim2.5$ at larger distances along the PWN tail. 

The excess map reported in Fig.~\ref{E_20TeV_skymap} at $E >20$~TeV traces emission from the vicinity of PSR\,J1826$-$1256. TeV photons could be produced by these synchrotron X-ray-emitting electrons through inverse Compton scattering off the CMB and IR photon fields. 
 For the latter, an energy density at a distance of 4 kpc $\sim$0.23~eV~cm$^{-3}$ was assumed (see e.g., \citealp{Vernetto2016}; we note, however, that the distance to PSR\,J1826$-$1256 is not very well constrained; e.g., \citealp{Wang2011} favored a distance of about 1.2 kpc, whereas \citealp{Karpova2019} found a distance of $\sim3.5$~kpc based on an empirical interstellar absorption-distance correlation). The CMB photon field provides $u_{\rm CMB}\sim $0.26~eV~cm$^{-3}$. Electron Lorentz factors of $\gamma_{\rm e}\sim1.3 \times 10^8$ and $\gamma_{\rm e}\sim3.6 \times 10^7$ would be required to reach the VHE $\gamma$-ray fluxes reported here for CMB and IR seed photons, respectively, in agreement with the estimates in \cite{Duvidovich2019}. 

The complex region around HESS\,J1826$-$130 has also been observed at TeV energies with the HAWC telescopes. A spatially coincident source, 2HWC\,J1825$-$13, is included in the 2$^{\rm nd}$ HAWC catalog \citep{2hwccat}. Furthermore, the HAWC collaboration recently reported the detection of $E > 100$~TeV emission from eHWC\,J1825$-$134 \citep{HAWC2019}. Its position and extension are marked with a white dotted circle in Fig.~\ref{E_20TeV_skymap}. With a half-width of $\sim0.36
^\circ$, the source encompasses a local peak of the $^{12}$CO distribution. It is also compatible with the location of the $E > 20$ TeV $\gamma$-rays from HESS\,J1825$-$137, and adjacent to HESS\,J1826$-$130 emission $E> 20$ TeV. On spectral grounds, both HESS\,J1825$-$137 and HESS\,J1826$-$130 display statistically preferred cut-offs and/or curvature in their spectral analysis (see Sect.~\ref{spect} and \citealp{hessj1825}, respectively). However they may still be able to partially contribute to the $\gtrsim 100$\,TeV flux toward eHWC J1825-134 \citep{HAWC2019}. In this regard, the preliminary results reported in \cite{Salesa2019} seem to support this scenario. 

\section{Conclusions}\label{conc}

Finally, HESS\,J1826$-$130 is a particle accelerator in the Galactic plane capable of delivering TeV emission above several tens of TeV. The deep observation campaign on the source reported in this paper provides several constraints to its spectral and morphological properties, although its origin is still uncertain. The presence of nearby SNRs found in the region and/or another yet unresolved accelerator, together with the gas density properties of the surroundings, could make hadronic interactions responsible for the observed emission. If protons are delivered with energies up to hundreds of TeV, HESS\,J1826$-$130 could point to the existence of a distinct population of CR sources that, along with the GC, significantly contribute to the Galactic CR flux around the knee feature. In a leptonic scenario, HESS\,J1826$-$130 has been proposed as the counterpart of the Eel Nebula powered by the pulsar PSR\,J1826$-$1256. Our observations reveal VHE emission above several tens of TeV that comes from the vicinity of the pulsar. Regarding the association of \J1826 with the recently discovered $\gtrsim 100$~TeV emitter, 2HWC\,J1825$-$13, it is worth noting that the spectrum reported in Fig.~\ref{spectrum} extends at least up to $\sim43$~TeV. Therefore, it could be possible that a fraction of the emission reported from eHWC\,J1825$-$134 could be partially contributed by HESS\,J1826$-$130 (see \citealp{Salesa2019}). On the other hand, HESS\,J1825$-$137 also displays VHE emission above tens of TeV, albeit displaying a softer spectrum. Observations of the region, for example using the LHAASO observatory (\citealp{Bai2019}), and in the future with the Cherenkov Telescope Array \citep{ctascience} with its superior angular and energy resolution and sensitivity at energies above 10 TeV, will provide further insights into the origin of \J1826.

\begin{acknowledgements}

The support of the Namibian authorities and of the University of Namibia in facilitating 
the construction and operation of H.E.S.S. is gratefully acknowledged, as is the support 
by the German Ministry for Education and Research (BMBF), the Max Planck Society, the 
German Research Foundation (DFG), the Helmholtz Association, the Alexander von Humboldt Foundation, 
the French Ministry of Higher Education, Research and Innovation, the Centre National de la 
Recherche Scientifique (CNRS/IN2P3 and CNRS/INSU), the Commissariat à l’énergie atomique 
et aux énergies alternatives (CEA), the U.K. Science and Technology Facilities Council (STFC), 
the Knut and Alice Wallenberg Foundation, the National Science Centre, Poland grant no. DEC-2017/27/B/ST9/02272, the South African Department of Science and Technology and National Research Foundation, the University of Namibia, the National Commission on Research, Science \& Technology of Namibia (NCRST), 
the Austrian Federal Ministry of Education, Science and Research and the Austrian Science Fund (FWF), 
the Australian Research Council (ARC), the Japan Society for the Promotion of Science and by the 
University of Amsterdam. We appreciate the excellent work of the technical support staff in Berlin, 
Zeuthen, Heidelberg, Palaiseau, Paris, Saclay, Tübingen and in Namibia in the construction and 
operation of the equipment. This work benefited from services provided by the H.E.S.S. 
Virtual Organisation, supported by the national resource providers of the EGI Federation. 

\end{acknowledgements}

\bibliographystyle{aa} 
\bibliography{HESSJ1826-130_An_extreme_particle_accelerator_in_the_Galactic_plane_arxiv} 
\end{document}